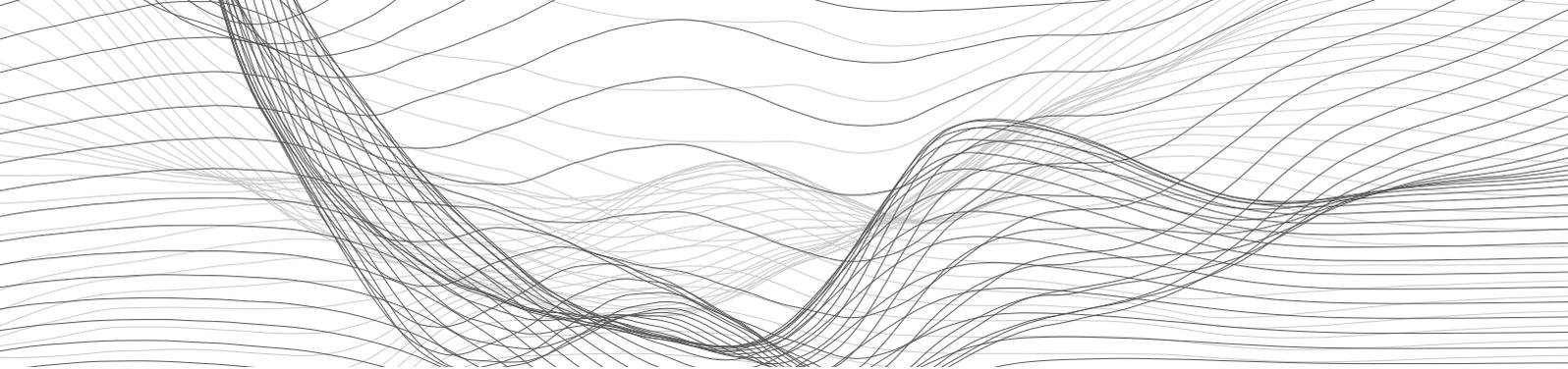

# How to issue a central bank digital currency

David Chaum, Christian Grothoff, Thomas Moser

SNB Working Papers
3/2021

# Legal Issues

# How to Issue a Central Bank Digital Currency[*]

*By* David Chaum [a], Christian Grothoff [b], and Thomas Moser [c]

[a] xx Network
[b] Bern University of Applied Sciences and GNU Project
[c] Swiss National Bank

This version: January 2021
First version: May 2020

**Abstract**

*With the emergence of Bitcoin and recently proposed stablecoins from BigTechs, such as Diem (formerly Libra), central banks face growing competition from private actors offering their own digital alternative to physical cash. We do not address the normative question whether a central bank should issue a central bank digital currency (CBDC) or not. Instead, we contribute to the current research debate by showing how a central bank could do so, if desired. We propose a token-based system without distributed ledger technology and show how earlier-deployed, software-only electronic cash can be improved upon to preserve transaction privacy, meet regulatory requirements in a compelling way, and offer a level of quantum-resistant protection against systemic privacy risk. Neither monetary policy nor financial stability would be materially affected because a CBDC with this design would replicate physical cash rather than bank deposits.*

[*] David Chaum (david@chaum.com), Christian Grothoff (christian.grothoff@bfh.ch), Thomas Moser (thomas.moser@snb.ch). We would like to thank Michael Barczay, Roman Baumann, Morten Bech, Nicolas Cuche, Florian Dold, Andreas Fuster, Stefan Kügel, Benjamin Müller, Dirk Niepelt, Oliver Sigrist, Richard Stallman, Andreas Wehrli, and three anonymous referees for comments and suggestions. The views, opinions, findings, and conclusions or recommendations expressed in this paper are strictly those of the author(s). They do not necessarily reflect the views of the Swiss National Bank (SNB). The SNB takes no responsibility for any errors or omissions in, or for the correctness of, the information contained in this paper.

# I. Introduction

Ever since the rise of personal computers in the 1980s, and especially since the National Science Foundation removed its restrictions on using the Internet for commercial purposes in 1991, there has been a quest to create digital cash for online payments. The earliest proposal was by Chaum (1983). Even though such schemes were implemented, they did not catch on; instead, credit card systems became the dominant online payment method. Nakamoto's (2008) proposal for a purely peer-to-peer version of digital cash and the subsequent successful launch of Bitcoin have unleashed a new era of digital cash research and development. CoinMarketCap lists over 5,000 cryptocurrencies. Recently, central banks have started to consider, or at least study, the issuance of digital currencies (see Auer et. al. 2020, Boar et al. 2020, Kiff et al. 2020, and Mancini-Griffoli et al. 2018).

Currently, central banks issue two types of money: (i) reserves in the form of settlement accounts at the central banks for selected financial market participants and (ii) currency in the form of banknotes available to the public. Accordingly, the literature on central bank digital currency (CBDC) distinguishes between (a) wholesale CBDC with limited access and (b) retail CBDC accessible to the public (see, e.g., Bech and Garratt 2017). A wholesale CBDC would be less disruptive to the current system since banks and selected financial market participants already have access to digital central bank money in the form of central bank accounts, which they use to settle interbank payments. Here, the question is whether the tokenization of central bank money and distributed ledger technology (DLT) offer distinct benefits over the existing real-time gross settlement (RTGS) systems. Thus far, the conclusion is that they do not, at least for domestic interbank payments (see Chapman et al. 2017).

A retail CBDC, which would be a new form of publicly available central bank money, could be more disruptive to the current system, depending on its design. The more such a CBDC would compete with commercial bank deposits, the greater the threat to bank funding with a potential adverse impact on bank credit and economic activity (see Agur et al. 2019). However, a retail CBDC could also have benefits (see Bordo and Levin 2017,

Berentsen and Schär 2018, Bindseil 2020, Niepelt 2020, Sveriges Riksbank 2020, and Bank of England 2020). Making electronic central bank money without counterparty risk available to everyone could improve the stability and resilience of the retail payment system. It could also provide a neutral payment infrastructure to promote competition, efficiency, and innovation. Overall, the costs and benefits of a retail CBDC are likely to differ from one country to another. For the view of the Swiss National Bank, which has no plans to issue a retail CBDC, see Jordan (2019).

The present paper focuses on retail CBDC, but we do not address whether a central bank *should* issue a CBDC. Instead, we focus on the potential design of CBDC. There has recently been increasing interest in CBDC design (see e.g. Allen et al. (2020), Bank of England (2020)). The design that we propose differs significantly from other proposals. Our system builds on and improves the eCash technology described by Chaum (1983) and Chaum et al. (1990). In particular, we propose a token-based, software-only CBDC without DLT. DLT is an interesting design if no central party is available or if the interacting entities are not willing to agree on a trusted central party. However, this is hardly the case for a retail CBDC issued by a *central* bank. Distributing the central bank's ledger with a blockchain merely increases transaction costs; it does not provide tangible benefits in a central bank deployment. Utilizing DLT to issue digital cash may be useful if there is no central bank to start with (e.g., the Marshall Islands' Sovereign project) or if the explicit intention is to do without a central bank (e.g., Bitcoin).[1]

The token-based CBDC proposed here also allows the preservation of a key feature of physical cash: transaction privacy. It is usually argued that cryptographic privacy protections are so computationally demanding that the high resource requirements make their use on mobile devices infeasible (see Allen et al. 2020). While this may be true in the context of DLT, where public traceability of transactions is necessary to prevent double-spending (Narayanan et al. 2016), it is not true for the Chaum-style blind-signature protocol with a central bank proposed in the present paper. Our CBDC, based on blind signatures and a two-tier architecture, guarantees perfect, quantum-resistant transaction privacy while providing anti-money laundering (AML) and counter terrorism financing (CFT) protections for society that are actually stronger than those of banknotes.

---

[1] There may be good use cases for DLT in the case of financial market infrastructure, such as digital exchanges, where the question of how to get central bank money onto the DLT for settlement purposes arises. However, in those situations, the potential benefits of DLT, e.g., lower costs or automatic reconciliation, do not arise from a decentralized issuance of central bank money.

Transaction privacy is important for three reasons. First, it protects users from government scrutiny and surveillance abuses. Mass surveillance programs are problematic even if people believe they have nothing to hide, simply because of the potential for error and abuse, particularly if such programs lack transparency and accountability (see Solove 2011). Second, transaction privacy protects users from data exploitation by payment service providers. Third, it protects users from the other party in a transaction, ruling out the possibility of ex-post opportunistic behavior or security risks due to the failure or neglect of customer data protection (see Kahn et al. 2005).

The paper is structured as follows: In Section II, we explain the difference between central bank money and other monies. In Section III, we review common, simplistic CBDC designs before proposing our design in Section IV. We then discuss regulatory and policy considerations (V) and related work (VI) and conclude (VII).

## II. What Is Central Bank Money?

Money is an asset that can be used to purchase goods and services. To be considered money, the asset must be accepted by entities other than the issuer. This is why vouchers, for instance, are not considered money. Genuine money must be *commonly* accepted as a medium of exchange. While money has other functions—such as being a unit of account and a store of value—its distinguishing feature is its function as a medium of exchange. Normally, the unit of account (i.e., how prices are quoted and debts are recorded) coincides with the medium of exchange for reasons of convenience. Separation can occur, however, if the medium of exchange lacks stability in value relative to the goods and services traded.[2] Money must also be a store of value to act as a medium of exchange because it has to preserve its purchasing power between the time it is received and spent. However, several other assets serve as stores of value, such as equities, bonds, precious metals, and real estate. Being a store of value is thus not a distinctive feature of money.

In a modern economy, the public uses two different types of money: (a) state money and (b) private money. State money is typically issued by the central bank, acting as an agent of the state. Central bank money is available to selected financial institutions in the form of deposits at the central bank (reserves) and to the public in the form of currency (i.e., banknotes and coins), also referred to as "cash." In a modern fiat money economy, such money has no intrinsic value. Legally, it is a liability of the central bank, although it is not redeemable. In most countries, central bank money is defined as legal tender, which means that it must be accepted toward repayment of a monetary debt, including taxes and legal fines. While this ensures that central bank money has some value, having legal-tender status is insufficient for central bank money to maintain a stable value. Rather, it is the central banks' monetary policy that maintains the money's value. Maintaining price stability—that is, a stable value of money relative to the value of the goods and services traded—is one of the central banks' main responsibilities.

Most payments in a modern economy are made with private money issued by commercial banks. Such money is composed of demand deposits that people have at commercial banks. These bank deposits can be accessed with checks, debit cards, credit cards, or other means of transferring money. They are a liability of the respective commercial bank. A key feature of bank deposits is that commercial banks guarantee convertibility on demand to central bank money at a fixed price, namely, at par. Depositors can withdraw their funds in cash or transfer the funds at a fixed rate of 1:1. Pegging their money to central bank money is how commercial banks maintain the value of their money.

---

[2] This can occur spontaneously in a high-inflation environment, e.g., when prices are quoted in USD but payments are made in local currency. The same is true for payments in Bitcoin, where prices are usually quoted in USD or other local currencies because of Bitcoin's high volatility. A separation can also occur by design, e.g., the Unidad de Fomento (UF) in Chile or the Special Drawing Right (SDR) of the International Monetary Fund (IMF). However, then, too, the purpose is to have a more stable unit of account.

Nevertheless, in a fractional reserve system, a commercial bank—even if solvent—may lack the liquidity to honor its promise to convert bank deposits to central bank money (e.g., in the case of a bank run) such that customers cannot withdraw their money. A bank can also become insolvent and go bankrupt, and customers can lose money as a result. Thus, commercial banks are regulated to mitigate such risks.

A significant difference between central bank money and privately issued commercial bank money is, therefore, that the latter entails counterparty risk. A central bank can always meet its obligations using its own nonredeemable money. Central bank money is the only monetary asset in a domestic economy without credit and liquidity risk. Therefore, it is typically the preferred asset to settle payments in financial market infrastructures (see, e.g., CPMI-IOSCO Principles for Financial Market Infrastructures (2012)). Another difference is that central bank money anchors the domestic monetary system by providing a reference of value with which private commercial bank monies maintain par convertibility.

Apart from commercial banks, other private entities occasionally attempt to issue money; cryptocurrencies are only the most recent attempt. But unlike bank deposits, such money is not commonly accepted as a medium of exchange. This is also true of Bitcoin, the most widely accepted cryptocurrency. One impediment to their usefulness as a medium of exchange is the high volatility of their value. A recent response to this problem was the emergence of stablecoins. Stablecoins generally attempt to stabilize their value in one of two ways: either by imitating central banks (algorithmic stablecoins) or by imitating commercial banks or investment vehicles (asset-backed stablecoins).[3]

"Algorithmic stablecoins" rely on algorithms to adjust their supply. In other words, they attempt to achieve price stability with their own "algorithmic monetary policy." There are examples of such stablecoins (e.g., Nubits), but so far, none has successfully stabilized its value over a long period of time.

---

[3] For a more detailed taxonomy and description of stablecoins, see Bullmann et al. (2019).

"Asset-backed" stablecoins differ according to the type of assets used and the legal rights that holders of stablecoins acquire. The types of assets typically used are money (central bank reserves, banknotes or commercial bank deposits), commodities (e.g., gold), securities, and sometimes other cryptocurrencies. How well such a scheme stabilizes the coins' value relative to the underlying asset(s) depends crucially on the legal rights that holders of the stablecoins acquire. If a stablecoin is redeemable at a fixed price (e.g., 1 coin = 1 USD, or 1 coin = 1 ounce of gold), such stability can theoretically be achieved.[4] What the scheme then essentially does is replicate commercial banks by guaranteeing convertibility into the underlying asset on demand. However, unlike bank deposits, which are typically only partially backed by central bank money reserves, stablecoins are generally fully backed by reserves of the underlying asset to avoid liquidity risk, mainly because they lack the benefits of public backstops such as deposit insurance and lender-of-last-resort support that apply to regulated banks.

Stablecoins backed by money are also called fiat-currency stablecoins. Holding 100 percent collateral in money (banknotes or bank deposits) is not very profitable, however. Accordingly, stablecoin providers have an incentive to economize their asset holdings and move to a fractional reserve system, just as commercial banks did.[5] This implies that they minimize their holdings of low-yielding assets to the minimum considered necessary to satisfy the convertibility requirement and add higher-yielding liquid assets such as government bonds instead. This improves their profitability but also increases their level of risk. However, even if a stablecoin is 100 percent collateralized by commercial bank deposits, it is still exposed to the credit and liquidity risks of the underlying bank. This risk can be removed if the deposits are held at the central bank so that the stablecoin is backed by central bank reserves. Such stablecoins have been called "synthetic CBDCs" (Adrian and Mancini-Griffoli 2019). It is important to point out, however, that such stablecoins are not central bank money and thus not CBDC since they are not a liability of the central bank and, therefore, are still subject to counterparty risk, namely, the risk that the issuer of the stablecoin goes bankrupt.

---

[4] Whether it also stabilizes the stablecoin's value relative to the goods and services traded depends on how stable the value of the respective asset is relative to the value of the goods and services.

[5] Uncertainty about whether a stablecoin is fully collateralized can be one of the reasons why a stablecoin can trade below par in the secondary market (see Lyons and Ganesh Viswanath-Natraj, 2020). This was historically also the case with banknotes when they were still issued by commercial banks. Such banknotes used to trade at varying discounts in the secondary market before the issuance of banknotes was nationalized and transferred to central banks as a monopoly.

If a stablecoin is not redeemable at a fixed price, its stability relative to the underlying asset is not guaranteed. If the stablecoin nevertheless represents an ownership share of the underlying asset, the scheme resembles a closed-end mutual fund or exchange-traded fund (ETF), and the corresponding risks apply. The value of the coin will depend on the net asset value of the fund, but its actual value can deviate. If there are authorized participants who can create and redeem stablecoins and thus act as arbitrageurs, as in the case of ETFs and as envisaged for Diem (Libra Association 2020), the deviation is likely to be small.

Overall, stablecoins have a greater chance of becoming money than do cryptocurrencies, especially if properly regulated. However, the availability of CBDCs would significantly limit their usefulness.

### III. Simplistic CBDC Designs

As noted, a CBDC would be a liability of the central bank. Two possible designs discussed in the literature are (a) an account-based CBDC and (b) a token-based (or valued-based) CBDC. These correspond to the two existing types of central bank money and corresponding payment systems (Kahn and Roberds 2008): central bank reserves (an account-based system) and banknotes (a token-based system). A payment occurs if a monetary asset is transferred from a payer to a payee. In an account-based system, a transfer occurs by charging the payer's account and crediting the payee's account. In a token-based system, the transfer occurs by transferring the value itself or a token, that is, an object that represents the monetary asset. The prime example of a token is cash—coins or banknotes. Paying with cash means handing over a coin or banknote. There is no need to record the transfer; possession of the token is sufficient. Therefore, parties are not required to reveal their identities at any time during the transaction; both can remain anonymous. However, the payee must be able to verify the token's authenticity. This is why central banks invest heavily in security features for banknotes.

It has been suggested that the distinction between account- and token-based systems is not applicable to digital currencies (Garratt et al. 2020). We are of a different opinion because we believe there is a significant difference. The critical distinction is the information carried by the information asset. In an account-based system, the assets (accounts) are associated with transaction histories that include all of the credit and debit operations involving the accounts. In a token-based system, the assets (tokens) carry information about their value and the entity that issued the token. The only possibility of attaining the transaction privacy property of cash, therefore, lies in token-based systems.[6]

*A. Account-Based CBDC*

The simplest way of launching a CBDC would be to allow the public to hold deposit accounts with the central bank. This implies that the central bank would be responsible for conducting know-your-customer (KYC) checks and ensuring AML/CFT compliance. This would include not only handling the initial KYC process but also authenticating customers for bank transactions, managing fraud, and dealing with false-positive and false-negative authentications. Given the limited physical presence of central banks in society and the fact that citizen authentication is currently not something central banks are likely prepared to do on a large scale, any account-based CBDC would require the central bank to outsource these checks. The entire servicing and maintenance of such accounts could be assigned to third-party providers (Bindseil 2020), or legislation could mandate commercial banks to open central bank accounts for their customers (Berentsen and Schär 2018).

---

[6] While the term "Bitcoin" suggests the use of a token, Bitcoin is an account-based system. The only difference between a traditional account-based system and a blockchain is that the accounts are not kept in a central database but in a decentralized append-only database.

Such an account-based CBDC would would potentially give a central bank a lot of information. One possible concern could be that this would allow governments to easily perform mass surveillance and levy sanctions against individual account holders. Their centralized nature makes such interventions inexpensive and easy to enforce against individuals or groups. Even in democracies, there are many examples of abuses of surveillance targeting critics and political opponents. One could argue that independent central banks could safeguard such information from government scrutiny and political abuse, but this would only open up a new avenue for political pressure, threatening central bank independence. In addition, the central database would be a significant target for attackers: even read-only access to parts of the database could create significant risks for people whose data might be exposed.

By providing bank accounts to the public, a central bank would also be in direct competition with commercial banks. This competition would entail two risks. First, it could threaten the deposit base of banks and, in the extreme, disintermediate the banking sector. This could adversely affect the availability of credit to the private sector and, as a result, economic activity (Agur et al. 2019). The disintermediation of banks could also lead to the centralization of the credit allocation process within the central bank, which would negatively affect productivity and economic growth. Second, allowing people to shift their deposits into a central bank safe haven could speed up bank runs during financial crises.

However, there are counterarguments. Brunnermeier and Niepelt (2019) argue that the transfer of funds from deposit to CBDC accounts would lead to an automatic substitution of deposit funding with central bank funding, merely rendering the central bank's implicit lender-of-last-resort guarantee explicit. Berentsen and Schär (2018) maintain that competition from central banks could even have a disciplinary effect on commercial banks and thus increase financial system stability, as commercial banks would have to make their business models more secure to avoid bank runs.

There are also proposals to mitigate the risk of disintermediation that aim at limiting or disincentivizing the use of CBDC as a store of value. One proposal is to cap the amount of CBDC that one can hold. A second proposal is to apply an adjustable interest rate to CBDC accounts so that the remuneration is always far enough below the remuneration at commercial bank accounts (possibly including a negative return) to make CBDC less attractive as a store of value (Kumhof and Noone 2018, Bindseil 2020). Further, to deter bank runs, Kumhof and Noone (2018) suggest that CBDC should not be issued against bank deposits but only against securities such as government bonds. Overall, an account-based CBDC would require further analysis of these issues.

### B. Hardware-dependent Token-Based CBDC

A central bank could also issue electronic tokens instead of accounts. Technically, this requires a system of ensuring that the electronic tokens are not copied easily. Physically unclonable functions (see Katzenbeisser et al. 2012) and secure zones in hardware (see Alves and Felton 2004, Pinto and Santos 2019) are two potential digital copy-prevention technologies. Physically unclonable functions, however, cannot be exchanged over the Internet (eliminating the main use case of CBDCs), and previous security features in copy-prevention hardware have been repeatedly compromised (see, e.g., Wojtczuk and Rutkowska 2009, Johnston 2010, Lapid and Wool 2019).

A key benefit of token-based CBDCs over central bank accounts is that token-based systems would work offline; that is, users could exchange tokens (peer-to-peer) without involving the central bank, which would protect individuals' privacy and liberty. However, the disintermediation that arises when users can trade electronic tokens without banks as intermediaries performing KYC checks and AML/CFT procedures would make it challenging to limit criminal abuse.

SIM cards are currently the most extensively available candidates for a secure hardware-based payment system, but they also have risks. Experience (e.g., see Soukup and Muff 2007, Garcia et. al. 2008, Kasper et. al. 2010, CCC 2017) suggests that any economically producible device that stores tokens with monetary value in an individual's possession, and enables offline transactions—and thus theft by cloning the information it contains—will be the target of successful forgery attacks as soon as the economic value from an attack would be sufficiently large. Such attacks include users who attack their own hardware (see also Allen et al. 2020). Payment card systems that have previously been deployed rely on tamper resistance in combination with fraud detection to limit the impact of a compromise. However, fraud detection requires the ability to identify payers and track customers, which is not compatible with transaction privacy.

## IV. A Token-Based CBDC Design to Guard Privacy

The CBDC proposed here is "software-only", merely a smartphone app that does not require any additional hardware from users. The CBDC builds on eCash and GNU Taler. Taler is part of the GNU Project, whose founder, Richard Stallman, coined the term "Free Software," currently often referred to as "Free/Libre and Open Source Software" (FLOSS).[7] Software is considered "Free Software" if its license grants users four essential freedoms: the freedom to run the program as they wish, the freedom to study the program and change it, the freedom to redistribute copies of the program, and the freedom to distribute copies of modified versions of the program. Free Software does not have to be noncommercial: providing software support is a standard business model for FLOSS.

Given the large numbers of stakeholders involved with a retail CBDC (the central bank, the financial sector, merchants, and customers) and the critical significance of the infrastructure, a retail CBDC should be based on FLOSS. Imposing a proprietary solution that requires dependence on a particular vendor would likely be an obstacle to adoption from the beginning. With FLOSS, all interested parties have access to every detail of the solution and the right to tailor the software to their needs. This leads to easier integration and better interoperability and competition among providers.[8] Moreover, it enables the central bank to meet the requirements for transparency and accountability. The benefits of FLOSS for security are also extensively recognized. The availability of the source code and the right to modify it make it easier to spot flaws and address them quickly.[9]

---

[7] For more information about GNU, see https://www.gnu.org and Stallman (1985). GNU Taler is released free of charge under the GNU Affero General Public License by the GNU Project. GNU projects popular among economists are the software packages «R» and "GNU Regression, Econometrics and Time-series Library" (GRETL). For a discussion of the benefits of FLOSS over proprietary software in research, see Baiocchi and Distaso (2003), Yalta and Lucchetti (2008), and Yalta and Yalta (2010). On open source licensing, see Lerner and Tirole (2005).

In our proposed architecture, all consumer and merchant interactions are with commercial banks. However, money creation and the database are provided exclusively by the central bank. The commercial banks authenticate the customers when they withdraw CBDC and the merchants/payees when they receive CBDC, but when spending CBDC, the customers/payers only need to authorize their transactions and do not need to identify themselves. This makes payments cheaper, easier, and faster, and it avoids easy interference with privacy (Dold 2019). In addition, authenticating customers when they withdraw CBDC and merchants/payees when they receive CBDC ensures KYC and AML/CFT compliance.

The CBDC proposed in the present paper is a genuine digital bearer instrument because when the user withdraws a sum of money in the form of a number, the number is "blinded" or hidden by the smartphone in a special encryption. In the actual system, a coin is a public/private key pair, with the private key only known to the owner of the coin.[10] The coin derives its financial value from the central bank's signature on the coin's public key. The central bank makes the signature with its private key, and it has multiple denomination key pairs available for blind-signing coins of different values. A merchant can use the central bank's corresponding "public key" to verify the signature. However, to be sure that the coin has not been copied and already redeemed by another payee (i.e., has not been "double-spent"), the merchant must deposit the coin so that the central bank can check the coin against a file of redeemed coins. Because neither the commercial bank nor the central bank see the coin's number during withdrawal, later, when the merchant deposits the coin, it is unknown which user withdrew it. The blinding and the resulting privacy are what make this type of CBDC a true digital bearer instrument.

---

[8] There may be some roles for private hardware, however. For example, protecting key stores and certain auditing functions, to the extent that such security can be shown to be only additive, may be an area where dedicated hardware evaluated by only a limited number of experts could have advantages.

[9] For instance, a cybersecurity bulletin issued by the U.S. National Security Agency in April 2020 urges users to prioritize open source software in the selection and use of collaboration services for Internet communication: "Open source development can provide accountability that code is written to secure programming best practices and isn't likely to introduce vulnerabilities or weaknesses that could put users and data at risk" (U/OO/134598-20).

[10] In Bitcoin, an account-based system, the key pair is an account, with the public key being the "address" of the account and thus a kind of "identity," even if pseudonymous.

In the subsequent discussion, we provide a high-level introduction to the technology and demonstrate how it can be integrated with the existing banking system to create a CBDC. Dold (2019) describes additional details.

*A. Key Building Blocks*

We now describe the main building blocks of the protocol, including the mathematical background for one possible instantiation of the cryptographic primitives used, to illustrate how an implementation could work. We note that alternative, equivalent mathematical designs exist for each component, and we are merely presenting the simplest secure designs of which we are aware of.

*Digital Signatures*—The basic idea of digital signatures in a public-key signature scheme is that the owner of a private key is the only one able to sign a message, while the public key enables anyone to verify the signature's validity.[11] The verification function's output is the binary statement "true" or "false". If the message is signed with the private key that belongs to the public verification key, the output is true; otherwise, it is false. In our proposal, the message is a "coin" or "banknote" with a serial number, and the central bank's signature affirms its validity. While GNU Taler by default uses modern EdDSA signatures (see Bernstein et al. 2012), we present a simple cryptographic signature scheme based on the well-studied RSA cryptosystem (Rivest et al. 1978).[12] However, in principle, any cryptographic signature scheme (DSA, ECDSA, EdDSA, RSA, etc.) can be used.

---

[11] Public-key cryptography was introduced by Diffie and Hellmann (1976), and the first implementation of digital signatures was introduced by Rivest, Shamir and Adleman (1978).

To generate RSA keys, a signer first picks two independent large primes $p$ and $q$ and computes $n = pq$ as well as Euler's totient function $\phi(n) = (p-1)(q-1)$. Then, any $e$ with $1 < e < \phi(n)$ and $gcd(e, \phi(n)) = 1$ can be used to define a public key $(e, n)$. The condition that the greatest common divisor (gcd) of $e$ and $\phi(n)$ has to be 1 (i.e., that they have to be relatively prime) ensures that the inverse of $e$ mod $\phi(n)$ exists. This inverse is the corresponding private key $d$. Given $\phi(n)$, the private key $d$ can be computed using the extended Euclidian algorithm such that $d \cdot e \equiv 1 \bmod \phi(n)$.

Given the private key $d$ and the public key $(e, n)$, a simple RSA signature $s$ over a message $m$ is $s \equiv m^d \bmod n$. To verify the signature, $m' \equiv s^e \bmod n$ is computed. If $m'$ and $m$ match, the signature is valid, which proves that the message was signed with the private key that belongs to the public verification key (message authentication) and that the message has not been changed in transit (message integrity). In practice, signatures are put over messages' hashes rather than the messages themselves. Hash functions compute digests of messages, which are short, unique identifiers for messages. Signing a short hash is much faster than signing a large message, and most signature algorithms only work on relatively short inputs.[13]

*Blind Signatures*—We use blind signatures, introduced by Chaum (1983), to protect the privacy of buyers. A blind signature is used to create a cryptographic signature for a message without the signer learning the contents of the message being signed. In our proposal, it prevents commercial banks and the central bank from tracing purchases back to the buyers. Our proposal works in principle with any blind signature scheme, but the best solution is still the RSA-based variant described by Chaum (1983).

---

[12] For a discussion of the RSA cryptosystem's long history and a survey of attacks on the RSA cryptosystem, see Boneh (1999).
[13] In the case of the RSA cryptosystem, the length limit is $log_2 n$ bits.

The blinding is done by the customers, who blind their coins before transmitting them to the central bank for signature. The customers therefore do not need to trust the central bank for privacy protection. Furthermore, RSA blinding would provide privacy protection even against quantum computer attacks. The central bank, for its part, sets up multiple-denomination key pairs available for blind-signing coins of different values, and it publishes/provides the corresponding public keys *(e, n)* for these values.

Let $f$ be the hash value of a coin and thus a unique identifier for the coin. The customer withdrawing the coin first generates a random blinding factor $b$ and computes $f' \equiv fb^e \mod n$ with the central bank's public key for that value. The blinded coin $f'$ is then transmitted to the central bank for signature. The central bank signs $f'$ with its private key $d$ by computing the blind signature $s' \equiv (f')^d \mod n$, appends the signature $s'$ to the blinded coin $f'$ and returns the pair $(s', f')$ to the customer. The customer can then unblind the signature by computing $s \equiv s'b^{-1} \mod n$. This works because $(f')^d = f^d b^{ed} = f^d b$ and, thus, multiplying $s'$ with $b^{-1}$ yields $f^d$, which is a valid RSA signature over $f$ as before: $s^e \equiv f^{de} \equiv f \mod n$.

In Chaum's original proposal, the coins were just tokens. However, we want consumers to be able to enter into contracts using digital signatures. To achieve this, whenever a digital wallet withdraws a coin, it first creates a random coin private key $c$ and computes the corresponding coin public key $C$ for creating digital signatures with regular cryptographic signature schemes (such as DSA, ECDSA, EdDSA, and RSA). Then, $f$ is derived using a cryptographic hash function from the public key $C$, which is then blindly signed by the central bank (using a fresh random blinding factor for each coin). Now, the customer can use $c$ to sign purchases electronically, thereby spending the coin.

As noted above, the central bank would establish key pairs for different coin values and publish the public keys that customers could use to withdraw money. These denomination keys, and thus the coins, would have an expiration date before which they must be spent or exchanged for new coins. Customers would be given a certain amount of time during which they could exchange their coins. A similar process exists for physical banknotes, where banknote series are regularly renewed so that the banknotes can be equipped with the latest security features, except that banknotes usually remain in circulation for decades rather than a few months or years.[14]

From a technical point of view, an expiration date has two advantages. First, it improves the efficiency of the system because the central bank can discard expired entries and does not have to store and search an ever-growing list of (spent) coins to detect double-spending. Second, it reduces security risks because the central bank does not have to worry about attacks against its expired (private) denomination keys ($d$). Moreover, even if a private key is compromised, the window during which the attacker can use the key is limited. In addition, charging an exchange fee would allow the central bank to implement negative interest rates, if deemed necessary. The central bank could also impose a conversion limit per customer for AML/CFT ("cash" limits) or financial stability reasons (to prevent hoarding or bank runs), if desired.

*Key-Exchange Protocol*—GNU Taler uses a key-exchange protocol in an unusual way to provide a link between an original coin and the change rendered for that original coin. This ensures that change can always be given without compromising income transparency or consumer privacy. The same mechanism can also be used to give anonymous customer refunds. The protocol also handles network and component failures, ensuring that payments either definitively succeeded or were definitively aborted and that all parties have cryptographic proof of the outcome. This is approximately equivalent to atomic swaps in interledger protocols or fair exchange in traditional e-cash systems.

The most common mathematical construction for a key-exchange protocol is the Diffie-Hellman construction (Diffie and Hellman 1976). It enables two parties to derive a shared secret key. To do this, they share two domain parameters $p$ and $g$, which can be public, where $p$ is a large prime number and $g$ is a primitive root modulo $p$.[15] Now, the two parties chose their private keys $a$ and $b$, which are two large integers. With these private keys and the domain parameters, they generate their respective public keys $A \equiv g^a \bmod p$ and $B \equiv g^b \bmod p$. Each party can now use its own private key and the other party's public key to compute the shared secret key $k \equiv (g^b)^a \equiv (g^a)^b \equiv g^{ab} \bmod p$.[16]

---

[14] In Switzerland, for instance, the Swiss National Bank began phasing out its eighth banknote series in April 2016. These banknotes were put into circulation in the late-1990s. Effective 1 January 2020, however, all banknotes starting from the sixth series issued in 1976 as well as any future series remain valid and can be exchanged for current notes indefinitely.

To obtain change, a customer starts with the private key of the partially spent coin $c$. Let $C$ be the corresponding public key, e.g., $C = g^c \bmod p$. When the coin was previously partially spent, the central bank recorded the transaction involving $C$ in its database. For simplicity, we will assume that a denomination exists that precisely matches this residual value. If not, the change protocol can simply be performed repeatedly until all of the change is obtained. Let $(e, n)$ be the denomination key for the change to be issued.

To obtain the change, the customer first creates $\kappa$ private transfer keys $t_i$ for $i \in \{1, \ldots, \kappa\}$ and computes the corresponding public keys $T_i$. These $\kappa$ transfer keys are simply public-private key pairs that allow the customer to run the key-exchange protocol locally—with the customer playing both sides—$\kappa$ times between $c$ and each $t_i$. If Diffie-Hellman is used for the key-exchange protocol, we will have $T_i \equiv g^{t_i} \bmod p$.

The result is three transfer secrets $K_i \equiv KX(c, t_i)$. The key-exchange protocol can be used in different ways to arrive at the same value $K_i \equiv KX(C, t_i) = KX(c, T_i)$. Given $K_i$, the customer uses a cryptographic hash function $H$ to derive values $(b_i, c_i) \equiv H(K_i)$, where $b_i$ is a valid blinding factor for the denomination key $(e, n)$ and $c_i$ is a private key for the fresh coin to be obtained as change. $c_i$ must be suitable for both creating cryptographic signatures and for future use with the key-exchange protocol (as $c$, to obtain change from the change). Let $C_i$ be the public key corresponding to $c_i$. The customer then requests the central bank to create a blind signature over $C_i$ for $\in \{1, \dots, \kappa\}$.[17] In this request, the customer also commits to the public keys $T_i$. The request is authorized using a signature made with the private key $c$.

---

[15] An integer $g$ is a primitive root modulo $p$ if for every integer $a$ coprime to $p$, there is some integer $k$ for which $g^k \equiv a \pmod{p}$. In practice, $g$ should be such a primitive $p$-$1$th root, which is also called a generator, in order to prevent subgroup attacks such as Pohlig-Hellman attacks (see Lim and Pil, 1997).

[16] The same mechanism could also be used to ensure that coins are not transferred to a third party during withdrawal. To achieve this, consumers would need to safeguard a long-term identity key. Then, the withdrawal process could use the same construction that GNU Taler uses to obtain change, except a customer's long-term identity key would be used instead of the original coin when withdrawing from the customer's bank account. However, a customer's failure to safeguard the long-term identity key could void the privacy assurances and enable the theft of all remaining coins. Given the limited risk in transfers to third parties when withdrawing coins, it is unclear whether this mitigation would be a good trade-off.

Instead of directly returning the blind signature, the central bank first challenges the customer to prove that the customer used the above construction correctly by providing a $\gamma \in \{1, \dots, \kappa\}$. The customer must then reveal the $t_i$ for $i \neq \gamma$ to the central bank. The central bank can then compute $K_i \equiv KX(C, t_i)$ and derive the $(b_i, c_i)$ values. If for all $i \neq \gamma$, the provided $t_i$ prove that the customer used the construction correctly, the central bank returns the blind signature over $C_\gamma$. If the customer fails to provide a correct proof, the residual value of the original coin is forfeited. This effectively punishes those trying to evade income transparency with an expected tax rate of $1 - \frac{1}{\kappa}$.

To prevent a customer from conspiring with a merchant who is trying to obscure its income, the central bank allows anyone who knows $C$ to, at any time, obtain the values of $T_\gamma$ and the associated blind signatures of all coins linked to the original coin $C$. This allows the owner of the original coin—who knows $c$—to compute $K_\gamma \equiv KX(c, T_\gamma)$ and, from there, to derive $(b_i, c_i)$ and, finally, to unblind the blind signature. Consequently, a merchant hiding income in this way would basically form a limited economic union with the customer instead of obtaining exclusive control.

## B. System Architecture

A key objective of our architecture is to ensure that central banks do not have to interact directly with customers or retain any information about them, merely maintaining a list of spent coins. The authentication is delegated to commercial banks that have the necessary infrastructure already in place. Withdrawal and deposit protocols reach the central bank via a commercial bank as an intermediary. From the customer standpoint, the process is analogous to withdrawing physical cash from an ATM. The transaction between a user's commercial bank and the central bank occurs in the background. Withdrawing CBDC would proceed as shown in Figure 1.

---

[17] If the RSA cryptosystem were to be used for blind signatures, we would use $f \equiv FDH_n(C_i)$, where $FDH_n(\ )$ is the full-domain hash over domain $n$.

**Figure 1. CBDC Withdrawal**

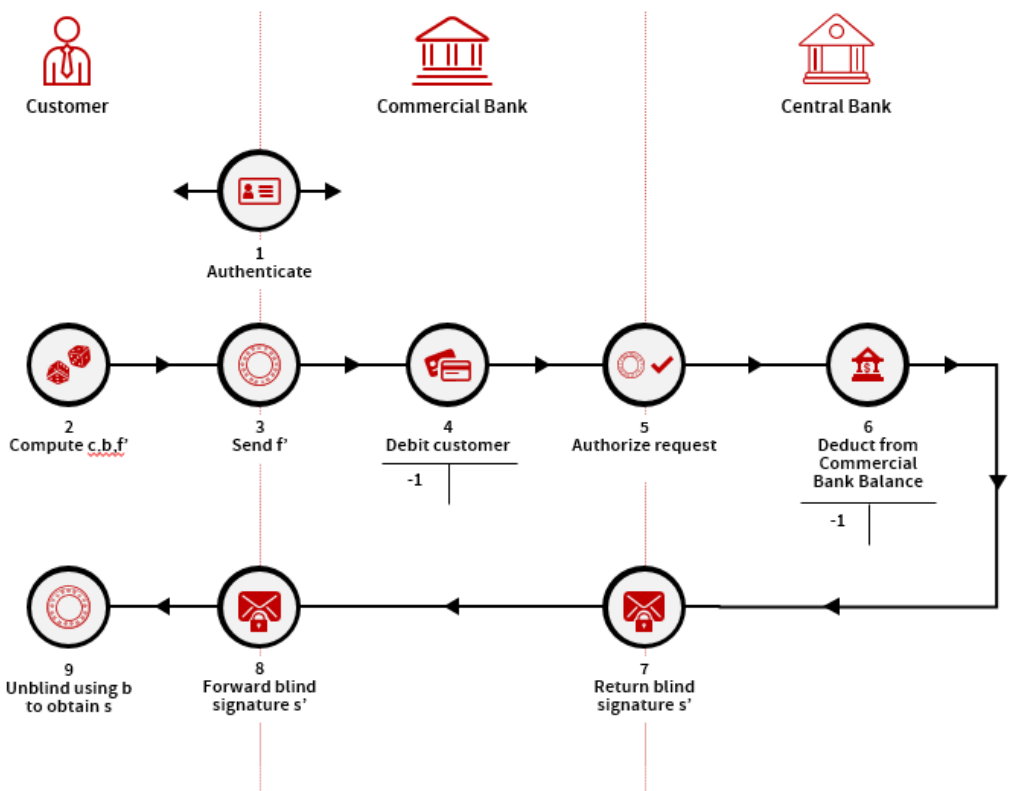

A customer (1) provides authentication to his or her commercial bank using the respective commercial bank's authentication and authorization procedures. Next, the customer's phone (or computer) obtains the public denomination key ($e$, $n$) provided by the central bank for that value; it then (2) computes a key pair for a coin, with a private key $c$ and a public key *C*, and chooses a blinding factor *b*. The coin's public key is then hashed (→ *f*) and blinded (→ *f′*). Next, (3) the customer's phone sends *f′* together with an authorization to withdraw the coin and debit the customer's account to the commercial bank via an established secure channel. The commercial bank then (4) debits the amount from the customer's deposit account, (5) digitally authorizes the request with its own bank-branch digital signature and forwards the request and the blinded coin to the central bank for signing. The central bank (6) deducts the value of the coin from the commercial bank's central bank account, blindly signs the coin with the central bank's private key for the respective value, and (7) returns the blind signature *s′* to the commercial bank. The commercial bank (8) forwards the blind signature *s′* to the customer's electronic wallet. Finally, the customer's phone (9) uses *b* to unblind the signature (→ *f*) and stores the newly minted coin (*c*, *s*).

When spending CBDC, the process is analogous to paying merchants in cash. However, to ensure settlement, the merchants must deposit the coins. Spending CBDC proceeds as shown in Figure 2.

A customer and a merchant negotiate a business contract, and (1) the customer uses an electronic coin to sign the contract/bill of sale with the coin's private key *c* and transmits the signature to the merchant. A coin's signature on a contract with a valid coin is an instruction from the customer to pay the merchant who is identified by the bank account in the contract. Customers may sign a contract with multiple coins if a single coin is insufficient to pay the total amount. The merchant then (2) validates the signature of the coin over the contract and the signature *s* of the central bank over *f* corresponding to the coin's *C* with the respective public keys and forwards the signed coin (together with the merchant's account information) to the merchant's commercial bank. The merchant's commercial bank (3) confirms that the merchant is one of its customers and forwards the signed coin to the central bank. The central bank (4) verifies the signatures and checks its database to ensure that the coin has not previously been spent. If everything is in order, (5) the central bank adds the coin to the list of spent coins, credits the commercial bank's account at the central bank and (6) sends a confirmation to the commercial bank to that effect. Next, (7) the commercial bank credits the merchant's account and (8) informs the merchant. The merchant (9) delivers the product/service to the customer. The entire process takes only a few hundred milliseconds.

**Figure 2. Spending and Depositing CBDC**

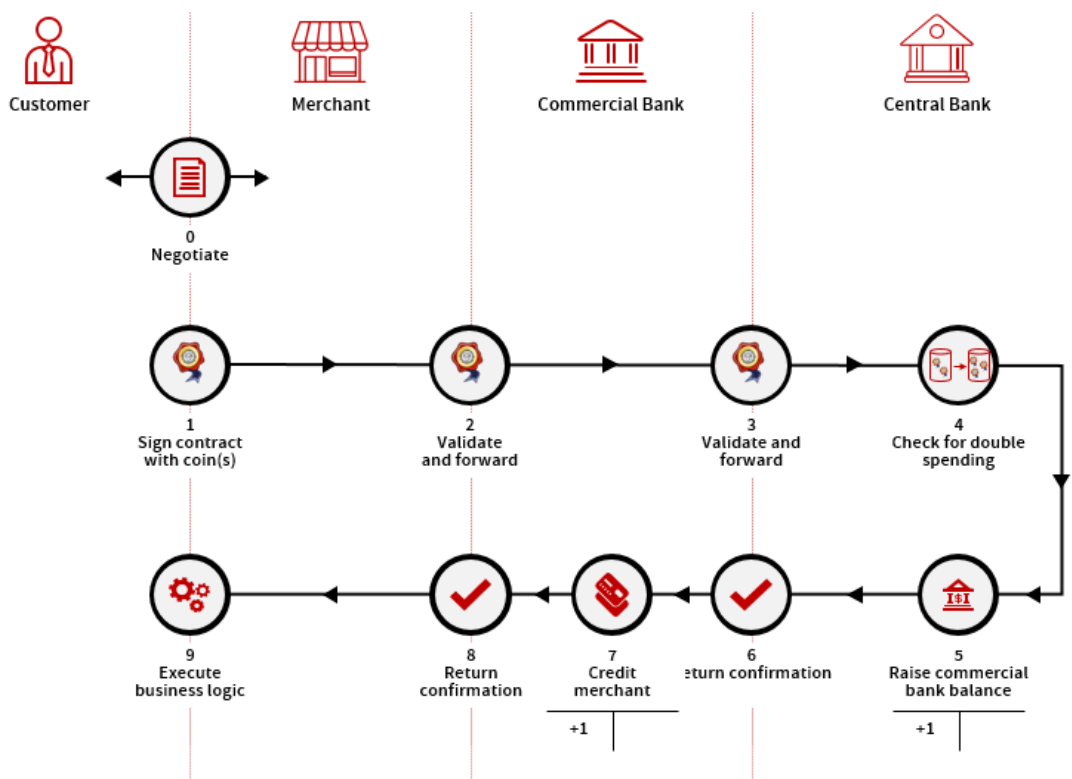

.

*C. Security Considerations*

Our proposal requires the central bank to operate a high-availability online service and database. Only online checks can effectively prevent double-spending because users can copy electronic coins. While theoretical solutions exist to retroactively identify users who engage in double-spending (see Chaum et al. 1990), these solutions create an economic risk for both users and the central bank, owing to the delayed identification of fraudulent transactions. Online double-spending detection eliminates this risk but in turn implies that transactions will be impossible if Internet connectivity to the central bank is unavailable.

The central bank will also need to protect the confidentiality of the private keys it uses to sign coins and other protocol messages. If the central bank's signing key(s) are ever compromised, such as by a quantum computer, by a physical attack on data-center vaults, or perhaps by some new unforeseen algorithm, the users can securely, and without compromising privacy, be refunded all the coins that they have not spent. The central bank would announce the key revocation via the Application Programming Interface (API), which would be detected by the wallets and cause them to initiate the following refresh protocol: the user discloses the coin's public key $C$, the central bank's signature $s$, and the blinding factor $b$ to the central bank, enabling the central bank to verify the user's legitimate withdrawal and to refund the unspent coin's value. To detect a possible compromise of its key, the central bank can monitor the database for instances of deposits exceeding withdrawals.

*D. Scalability and Costs*

Our proposed scheme would be as efficient and as cost-effective as modern RTGS systems currently used by central banks.

Scalability concerns the cost of growing processing capacity so that an increasing number of transactions can be processed with an adequate time to finality. The overall system cost can be low since the CBDC proposed here is software only. Spent coins must be stored until the denomination key pair that was used to sign the coins expires, for instance, via a rolling annual schedule, which keeps the database size bounded. The amount of additional processing power and bandwidth needed grows by the same amount for each additional transaction, spend or deposit, because the transactions are essentially independent of each other. This additional power is attained by simply adding more hardware, often called partitioning or sharding. With so-called consistent hashing, hardware additions need not be disruptive. Any underlying database technology can be utilized.

More concretely, the front-end logic at the central bank only needs to conduct a few signature operations, and a single CPU core can do a few thousand operations per second (see Bernstein and Lange 2020). If a single system is insufficient, it is easy to deploy additional front-end servers and direct the various commercial banks to balance their requests across the server farm or use a load balancer to distribute requests within the central bank's infrastructure.

The front-end servers need to communicate with a database to make transactions and to prevent double-spending. A single modern database server should be able to reliably handle tens of thousands of such operations per second. Operations are split easily across multiple database servers by merely assigning each database server a range of values for which it is responsible. This design ensures that individual transactions never cross shards. Hence, the back-end systems are also expected to scale linearly with the computing resources made available, again starting from a high baseline for a single system.

The front-ends also need to communicate with the back-ends using an interconnect. Interconnects can support very large numbers of transactions per second. The size of an individual transaction is typically approximately 1–10 kilobytes. Thus, modern datacenter interconnects, switching at rates of 400 Gbit/s, can support millions of transactions per second.

Finally, the total system cost is low. Securely storing 1–10 kilobytes per transaction for many years is likely to be the predominant cost of the system. Using Amazon Web Services pricing, we experimented with an earlier prototype of GNU Taler and found that the cost of the system (storage, bandwidth, and computation) at scale would be below USD 0.0001 per transaction (for details regarding the data, see Dold 2019).

## V. Regulatory and Policy Considerations

In the proposed scheme, central banks do not learn the identities of consumers or merchants or the total transaction amounts. Central banks only see when electronic coins are released and when they are redeemed. Commercial banks continue to provide crucial customer and merchant authentication and, in particular, remain the guardians of KYC information. Commercial banks observe when merchants receive funds and can limit the amount of CBDC per transaction that an individual merchant may receive, if required. Additionally, transactions are associated with the relevant customer contracts. The resulting income transparency enables the system to be compliant with the AML/CFT regulations. If unusual patterns of merchant income are detected, the commercial bank, tax authorities, or law enforcement can obtain and inspect the business contracts underlying the payments to determine whether the suspicious activity is nefarious. The resulting income transparency is also a strong measure against tax evasion because merchants cannot underreport income or evade sales taxes. Overall, the system implements privacy-by-design and privacy-by-default approaches (as, for instance, required by the EU's General Data Protection Regulation). Merchants do not inherently learn the identity of their customers, banks have only necessary insights into their own customers' activities, and central banks are blissfully divorced from detailed knowledge of citizens' activities.

In some countries, there are limits on cash withdrawals and payments for regulatory reasons. Such restrictions could also be implemented for CBDC in the proposed design. For example, consumers could be limited to withdrawing a certain amount per day, or commercial banks could be limited in terms of the total amount of CBDC they can convert.

A potential financial stability concern often raised with retail CBDCs is banking sector disintermediation. Notably, retail CBDCs could make it easier to hoard vast amounts of central bank money. This could adversely affect the deposit funding of banks because the public would hold less of its money in the form of bank deposits. For countries whose currencies serve as safe-haven currencies, it could further lead to increased capital inflows during global risk-off periods, resulting in additional exchange rate appreciation pressures.

While this would be a serious concern with an account-based CBDC, it should be less of a concern with a token-based CBDC. First, hoarding a token-based CBDC entails risks of theft or loss similar to those of hoarding cash. Holding a few hundred dollars on a smartphone is probably an acceptable risk for many, but holding a very large amount is probably a less acceptable risk. Therefore, we would not expect significantly more hoarding than in the case of physical cash.

However, should hoarding or massive conversions of money from bank deposits to CBDC become a problem, central banks would have several options. As noted, under the proposed design, central banks configure an expiration date for all signing keys, which implies that at a set date, the coins signed by those keys become invalid. When denomination keys expire and customers have to exchange coins signed with old denomination keys for new coins, the regulator could easily impose a conversion limit per customer to enforce a hard limit on the amount of CBDC that any individual can hoard. In addition, central banks could charge a fee if necessary. Such a refresh fee when coins are set to expire would effectively mean negative interest rates on CBDC and make CBDC less attractive as a store of value, as suggested by Bindseil (2020). In fact, it would be a straightforward implementation of Silvio Gesell's idea of a carry tax on currency, famously referenced by Keynes (1936) and revived by Goodfriend (2000), Buiter and Panigirtzoglou (2003), and Agarwal and Kimball (2019).

Concerning potential implications for monetary policy, we do not anticipate material effects because our CBDC is designed to replicate cash rather than bank deposits. The issuance, withdrawal and deposit of our CBCD correspond exactly to the issuance, withdrawal and deposit of banknotes. It is possible that a retail CBDC could have a different velocity of circulation than physical cash, but this would not be a material problem for monetary policy.

## VI. Related Work

As noted earlier, the CBDC proposed in the present paper is based on eCash and GNU Taler.[18] Since Chaum's original e-cash proposal, research has focused on three main issues. First, in Chaum's original proposal, the coins were of fixed value and could only be spent in their entirety. Paying vast amounts with coins denominated in cents would be inefficient, so Okamoto (1995), Camenisch (2005), Canard and Gouget (2007), and Dold (2019) invented ways to address this issue. These solutions involve protocols for giving change or for providing divisibility of coins.

A second issue is that transactions sometimes fail due to network outages, for instance. In this case, the system must make it possible for the funds to remain with the consumer without a negative impact on privacy. Endorsed e-cash proposed by Camenisch et al. (2007) and Dold (2019) both tackle this issue. Several of the above solutions violate the assurances of privacy for customers using these features, and all of them except Taler violate the requirement of income transparency.

The third major issue, often neglected, is maintaining income transparency and thus AML and KYC compliance. Fuchsbauer et al. (2009) deliberately designed their system for disintermediation to provide more cash-like semantics. However, unlimited disintermediation is typically at odds with AML and KYC regulations, as it becomes impossible to attain any level of accountability. An example of such a design is ZCash, a distributed ledger that hides payer, payee, and transaction amount information from the network and is thus the perfect payment system for online crime. Only Taler offers both consistent customer privacy and income transparency while also providing efficient change, atomic swaps (see Camenisch 2007), and the ability to restore wallets from backup.

Regarding payment systems for CBDCs, Danezis and Meiklejohn (2016) designed a scalable ledger with RSCoin. It is basically an RTGS system that is secured using the same cryptography that is used in Bitcoin. Like Taler, the design uses database sharding to achieve linear scalability. However, Danezis and Meiklejohn's design does not have any provisions for privacy and lacks considerations for how to practically integrate the design with existing banking systems and processes.

---

[18] The actual implementation of eCash by the DigiCash company in the 1990s is documented at https://www.chaum.com/ecash.

The European Central Bank's EUROchain (see ECB 2019) is another prototype for a CBDC with a distributed ledger. Similar to the architecture proposed in the present paper, EUROchain uses a two-tier architecture with commercial banks acting as intermediaries. One crucial difference is how the systems try to combine privacy and AML compliance. While in our design, regulators could impose a limit on the amount of electronic cash that a bank account holder can withdraw over a certain time, EUROchain issues a limited number of "anonymity vouchers" that grant the receiver a limited number of transactions without AML checks. As these vouchers seem divorced from any token of value, it remains unclear how the design could avoid the emergence of a black market for "anonymity vouchers". Moreover, EUROchain's notion of anonymity is very different, in that their "anonymity vouchers" merely eliminate certain AML checks while preserving commercial banks' ability to view how their consumers spend their electronic cash. Whereas Taler payers directly interact with merchants to spend their e-cash, the EUROchain system requires payers to instruct their commercial banks to access their CBDC. Therefore, EUROchain does not directly issue valuable tokens to consumers and instead relies on consumers to authenticate themselves to their commercial banks to access CBDC that the central bank effectively holds in escrow. Thus, it is unclear what privacy, performance, or security benefits EUROchain has over existing deposit money.

## VII. Conclusion

With the emergence of Bitcoin and the recently proposed digital currencies from BigTech, such as Diem (formerly Libra), central banks ace growing competition from actors offering their own digital alternative to physical cash. Central banks' decisions on whether to issue a CBDC or not depend on how they assess the benefits and risks of a CBDC. These benefits and risks, as well as the specific jurisdictional circumstances that define the scope of retail CBDC, are likely to differ from one country to another.

If a central bank decides to issue a retail CBDC, we propose a token-based CBDC that combines transaction privacy with KYC and AML/CFT compliance. Such a CBDC would not compete with commercial bank deposits but rather replicate physical cash, thereby limiting financial stability and monetary policy risks.

We have shown that the scheme proposed here would be as efficient and as cost-effective as modern RTGSs operated by central banks. Electronic payments with our CBDC would only necessitate simple database transactions and miniscule amounts of bandwidth. The efficiency and cost-effectiveness, together with enhanced consumer usability caused by shifting from authentication to authorization, make this scheme likely to be the first to support the long-envisioned objective of online micropayments. In addition, the use of coins to cryptographically sign electronic contracts would enable the use of smart contracts. This, too, could lead to the emergence of entirely new applications for payment systems. Although our system is not based on DLT, it could easily be integrated with such technologies if required by financial market infrastructures in the future.

Just as important, however, a retail CBDC must preserve cash as a privacy-friendly commons under citizens' individual control. This can be attained with the scheme proposed in this paper, and central banks can avoid significant disruptions to their monetary policy and financial stability while reaping the benefits of going digital.

# Recent SNB Working Papers

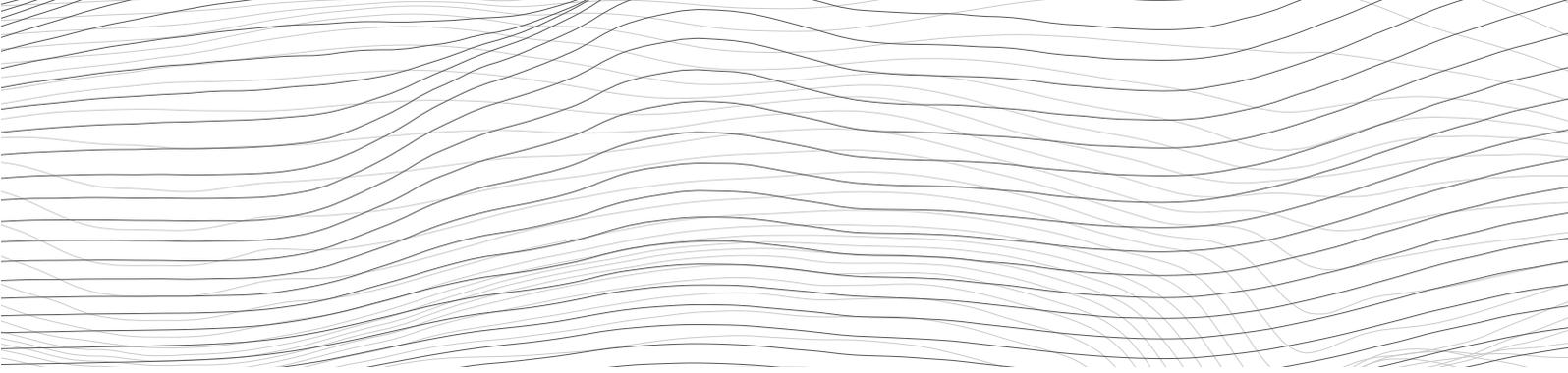